\documentclass[reprint,amsmath,amssymb,nofootinbib,superscriptaddress]{revtex4-1}

\usepackage{graphicx,hyperref, tabularx}

\usepackage{color}

\newcommand{\betamass}{$\beta_{i}=\{0.3,0,-0.3\}$}
\newcommand{\gammamass}{$\gamma^M_{\mathrm{sph}}=0.3753(1)$,
  $\gamma^M_{\Re(Y^2_1)}=0.3748(2)$, and
  $\gamma^M_{\text{3-d}}=0.3761(3)$}
\newcommand{\Deltamass}{$2\gamma^M_{\mathrm{sph}} T^M_{\mathrm{sph}}=3.46\pm 0.01$,
  $2\gamma^M_{\Re(Y^2_1)}T^M_{\Re(Y^2_1)}=3.46\pm 0.02$, and
  $2\gamma^M_{\text{3-d}}T^M_{\text{3-d}}=3.67\pm 0.04$}
\newcommand{\gammamassbootstrap}{$\gamma^M_{\mathrm{sph}}=0.3753(2)$,
  $\gamma^M_{\Re(Y^2_1)}=0.3750(5)$, and
  $\gamma^M_{\text{3-d}}=0.376(1)$}
\newcommand{\Deltamassbootstrap}{$2\gamma^M_{\mathrm{sph}} T^M_{\mathrm{sph}}=3.46\pm 0.03$,
  $2\gamma^M_{\Re(Y^2_1)} T^M_{\Re(Y^2_1)}=3.47\pm 0.06$, and
  $2\gamma^M_{\text{3-d}}T^M_{\text{3-d}}=3.7\pm 0.8$}

\newcommand{\betaricci}{$\beta_{i}=\{0.4, 0, -0.4\}$}
\newcommand{\gammaricci}{$\gamma^R_{\mathrm{sph}}=0.3787(1)$,
  $\gamma^R_{\Re (Y^2_1)}=0.3761(1)$, and $\gamma^R_{\text{3-d}}=0.3755(2)$}
\newcommand{\gammariccibootstrap}{$\gamma^R_{\mathrm{sph}}=0.379(3)$,
  $\gamma^R_{\Re (Y^2_1)}=0.376(3)$, and $\gamma^R_{\text{3-d}}=0.375(5)$}
\newcommand{\Deltariccibootstrap}{$2\gamma^R_{\mathrm{sph}}T^R_{\mathrm{sph}}=3.56\pm 0.02$,
  $2\gamma^R_{\Re(Y^2_1)}T^R_{\Re(Y^2_1)}=3.52\pm 0.05$, and
  $2\gamma^R_{\text{3-d}}T^R_{\text{3-d}}=3.51\pm 0.07$}

\begin{document}

\title{Critical behavior in 3-d gravitational collapse of massless scalar fields}

\author{Nils Deppe}
\email{nd357@cornell.edu}
\affiliation{Cornell Center for Astrophysics and Planetary Science,
  Cornell University, Ithaca, New York 14853, USA}
\author{Lawrence E.~Kidder}
\affiliation{Cornell Center for Astrophysics and Planetary Science,
  Cornell University, Ithaca, New York 14853, USA}
\author{Mark A.~Scheel}
\affiliation{Theoretical Astrophysics 350-17,
  California Institute of Technology, Pasadena, CA 91125, USA}
\author{Saul A.~Teukolsky}
\affiliation{Cornell Center for Astrophysics and Planetary Science,
  Cornell University, Ithaca, New York 14853, USA}
\affiliation{Theoretical Astrophysics 350-17,
  California Institute of Technology, Pasadena, CA 91125, USA}
\date{\today}

\begin{abstract}
  We present results from a study of critical behavior in 3-d gravitational
  collapse with no symmetry assumptions. The source of the gravitational field
  is a massless scalar field. This is a well-studied case for spherically
  symmetric gravitational collapse, allowing us to understand the reliability
  and accuracy of the simulations. We study both supercritical and subcritical
  evolutions to see if one provides more accurate results than the other.  We
  find that even for non-spherical initial data with 35 percent of the power in
  the $\ell=2$ spherical harmonic, the critical solution is the same as in
  spherical symmetry.
\end{abstract}

\maketitle

\section{Introduction}

Critical behavior in the gravitational collapse of a massless scalar field was
discovered by Choptuik~\cite{Choptuik:1992jv}, who sought to answer the question
``What happens at the threshold of black hole formation?''  Choptuik considered
a massless scalar field undergoing gravitational collapse in a spherically
symmetric spacetime. He found that for some parameter $p$ in the initial data,
for example the amplitude of a Gaussian-distributed scalar field, the final mass
of the black hole is related to $p$ by
\begin{align}
  \label{eq:simple mass relation}
  M_{\mathrm{BH}}\propto\left|\frac{p}{p_\star}-1\right|^{\gamma_M}.
\end{align}
Here $p_\star$ is the critical value of the parameter $p$ that separates initial
data that form a black hole (supercritical) from initial data that do not form a
black hole (subcritical). Choptuik observed that the critical exponent
$\gamma_M$ is independent of the initial data chosen---the critical behavior is
universal. The currently accepted value of the critical exponent is
$\gamma_{M}=0.374\pm0.001$~\cite{Gundlach:1996eg}. Not much later, Garfinkle and
Duncan~\cite{Garfinkle:1998va} discovered that in subcritical evolutions the
maximum absolute value of the Ricci scalar at the center of the collapse obeys
the scaling relation
\begin{align}
  \label{eq:simple ricci relation}
  R_{\max} \propto \left|\frac{p}{p_\star}-1\right|^{2 \gamma_{R_{\max}}}.
\end{align}
Interestingly, $\gamma_{R_{\max}}$ was found to have the same value as
$\gamma_M$.

Another key aspect of the critical behavior observed by Choptuik is that of a
discretely self-similar solution, or ``echoing''. In the strong-field regime
near the critical solution, Choptuik noticed that any gauge-invariant quantity
$U$ obeys the relation
\begin{align}
  \label{eq:rescaling}
  U(\mathcal{T}, x^i) = U(e^\Delta \mathcal{T}, e^\Delta x^i),
\end{align}
where $\Delta$ is a dimensionless constant.  Here $\mathcal{T}=\tau-\tau_\star$,
where $\tau$ is the proper time of a central observer and $\tau_\star$ is the
value of $\tau$ when a naked singularity forms in the limit $p\to p_\star$.
$\tau_\star$ is referred to as the accumulation time. As one
moves closer in time to the critical solution by $e^\Delta$, the same field
profile is observed for $U$ but at spatial scales $e^\Delta$ smaller. The
echoing period $\Delta$, like the critical exponent, is universal in the sense
that it does not depend on the initial data, only on the type of matter
undergoing gravitational collapse. The currently accepted value for a massless
scalar field is $\Delta=3.4453\pm0.0005$~\cite{Gundlach:1996eg}.

Since the seminal work by Choptuik, many studies to better understand critical
behavior in gravitational collapse have been performed.  Studies of critical
collapse of a massless scalar field in spherical symmetry have found that the
critical exponent and echoing period are both independent of the initial data
profile but depend on the dimensionality of the
spacetime~\cite{Garfinkle:1999zy,Bland:2005vu,Sorkin:2005vz,Taves:2011yt}. Similar
studies observed that the critical exponent, echoing period, and possibly even
the type of phase transition are changed in modified theories of
gravity~\cite{Deppe:2012wk,Golod:2012yt}.  Interestingly, the presence of
critical behavior appears to be independent of the matter source, but the value
of the critical exponent, echoing period, and type of phase transition depend on
the type of
matter~\cite{Choptuik:1996yg,Gundlach:1996je,Brady:1997fj,Garfinkle:2003jf,Baumgarte:2015aza,Gundlach:2016jzm,Baumgarte:2016xjw,Gundlach:2017tqq}.
Vacuum critical collapse was first studied
in~\cite{Abrahams:1993wa,Abrahams:1994nw}, which found that critical behavior is
present and that the critical exponent and echoing period have values different
from those found in simulations with matter.  Unfortunately, studying vacuum
gravitational collapse has proven to be quite
difficult~\cite{Sorkin:2009wh,Sorkin:2010tm,Hilditch:2013cba,Hilditch:2015aba}.

In critical collapse the phase transition is either Type I or Type II.
In Type II phase transitions the black hole mass continuously goes to zero as
$p_\star$ is approached. This has been the most common case observed so far when
studying critical collapse. In Type I transitions the mass of the black hole
that forms approaches a constant, non-zero value as $p_\star$ is
approached. Type I phase transitions have been clearly identified in critical
collapse of a massive scalar field\cite{Brady:1997fj}.
The discussion in this paper is only relevant for Type II critical behavior.

In 1997 both Gundlach~\cite{Gundlach:1996eg}, and Hod and
Piran~\cite{Hod:1996az} independently discovered fine structure in addition to
the power-law behavior of the black hole masses: There is a small-amplitude
modulation of~\eqref{eq:simple mass relation}. Specifically, the scaling
relation is altered to
\begin{align}
  \ln(M_{\mathrm{BH}})=&\gamma_{M}\ln\left|p/p_\star-1\right|+C\notag\\
                       &+A\sin(w\ln\left|p/p_\star-1\right|+\delta),
\end{align}
where $C$, $A$, $w$, and $\delta$ are constants.  These authors predicted and
verified that $w=\Delta/(2\gamma_{M})$ for massless scalar field collapse in
spherical symmetry. Whether or not this relation holds for different matter
sources and beyond spherical symmetry is an open question.

Unfortunately, answering the question of how symmetry assumptions affect the
critical exponent and echoing period has turned out to be quite challenging. The
reason is that spatiotemporal scales varying over four to six orders of
magnitude must be resolved in order to properly study the fine structure and
echoing, and a large number of high-resolution simulations are necessary. In
addition, the well-posedness and stability of the formulation of the Einstein
equations solved and the choice of gauge has proven to be as problematic here as
in other simulations in numerical relativity.  Akbarian and
Choptuik~\cite{Akbarian:2015oaa} have recently studied how formulations of the
Einstein equations commonly used for binary black hole mergers behave when
studying critical collapse. However, that work was restricted to spherical
symmetry.

Critical collapse of a massless scalar field in axial symmetry was studied using
perturbation theory by Martin-Garcia and Gundlach~\cite{MartinGarcia:1998sk},
who found that all non-spherical modes decay. In 2003 Choptuik
et.~al~\cite{Choptuik:2003ac} performed numerical simulations of massless scalar
field collapse in axial symmetry. They found that the critical solution in this
case is the same as the solution found in spherical symmetry.  However, in
contrast to~\cite{MartinGarcia:1998sk}, they also found tentative evidence for a
non-decaying $l=2$ mode. More recently, Healy and Laguna~\cite{Healy:2013xia}
studied critical collapse of a massless scalar field that is symmetric about the
$xz$-plane. Healy and Laguna observed results consist with spherically symmetric
collapse, but were unable to verify the echoing of gauge-independent fields. The
work of Healy and Laguna has been followed by a study of massless scalar field
collapse with a quartic potential by Clough and
Lim~\cite{Clough:2016jmh}. Clough and Lim also studied initial data similar to
that of~\cite{Healy:2013xia} and obtained results similar to those of Healy and
Laguna.

In this paper we present a study of critical collapse of a massless scalar field
with no symmetry assumptions, and the first study beyond spherical symmetry that
is able to resolve the fine structure in the black hole mass scaling
relation. We are able to resolve small-scale dynamics in both supercritical and
subcritical evolutions, allowing us to directly compare the results. In
$\S$\ref{sec:Equations} we review the equations solved, in
$\S$\ref{sec:InitialData} we discuss the initial data used, in
$\S$\ref{sec:NumericalMethods} we provide details about the numerical method, in
$\S$\ref{sec:Results} we present the results, and we conclude in
$\S$\ref{sec:Conclusions}.

After this work was completed, a paper by
Baumgarte appeared\cite{Baumgarte:2018fev} in which axially symmetric initial
data similar to that of~\cite{Choptuik:2003ac} is studied. We discuss the
relation between this paper and our work at the end of $\S$\ref{sec:Results}.

\section{Equations}\label{sec:Equations}

We study the dynamics near the critical solution in gravitational collapse of
the Einstein-Klein-Gordon system. We solve the Einstein equations,
\begin{align}
  \label{eq:EE}
  R_{ab}=8\pi\left(T_{ab}-\frac{1}{2}\psi_{ab}T^c{}_c\right)
\end{align}
where $R_{ab}$ is the Ricci tensor, $\psi_{ab}$ the spacetime metric, and
$T_{ab}$ the stress tensor. Here and throughout the rest of the paper we will
use latin indices at the beginning of the alphabet, e.g.~$a,b,c,\ldots$ to refer
to spacetime indices running from 0 to 3, and later indices, $i,j,k,\ldots$ to
refer to spatial indices running from 1 to 3. We use the ADM form of the metric,
\begin{align}
  ds^2=-N^2dt^2+g_{ij}\left(N^i dt + dx^i\right) \left(N^j dt + dx^j\right)
\end{align}
where $N(t,x^i)$ is the lapse, $N^j(t,x^i)$ the shift, and $g_{ij}(t, x^k)$ the
spatial metric. We denote the timelike unit normal orthogonal to the spacelike
hypersurfaces by
\begin{align}
  t^a = (N^{-1},-N^i/N).
\end{align}
We solve Eq.~(\ref{eq:EE}) using a first-order generalized harmonic (GH)
formulation~\cite{Lindblom:2005qh}.

The matter source is a massless scalar field $\varphi$ with
\begin{align}
  \label{eq:StressTensor}
  T_{ab}=\partial_a\varphi\partial_b\varphi-
  \frac{1}{2}\psi_{ab}\psi^{cd}\partial_c\varphi\partial_d\varphi.
\end{align}
To bring the resulting equations of motion into first-order form, we define the
auxiliary variables $\Phi_i=\partial_i\varphi$ and
$\Phi_{iab}=\partial_i\psi_{ab}$, and the conjugate variables
$\Pi=-N^{-1}\left(\partial_t\varphi-N^i\partial_i\varphi\right)$ and
$\Pi_{ab}=-N^{-1}\left(\partial_t \psi_{ab}-N^i\Phi_{iab}\right)$.

The first-order GH system is~\cite{Lindblom:2005qh}

\begin{align}
  \label{eq:metric_evolution}
  \partial_t\psi_{ab}-&\left(1+\gamma_1\right)N^k\partial_k\psi_{ab}=-N\Pi_{ab}-\gamma_1N^i\Phi_{iab},\\
  \label{eq:metric_conjugate_evolution}
  \partial_t\Pi_{ab}-&N^k\partial_k\Pi_{ab}+Ng^{ki}\partial_k\Phi_{iab}-\gamma_1\gamma_2N^k\partial_k\psi_{ab}\notag\\
  =&2N\psi^{cd}\left(g^{ij}\Phi_{ica}\Phi_{jdb}-\Pi_{ca}\Pi_{db}-\psi^{ef}\Gamma_{ace}\Gamma_{bdf}\right)\notag\\
                      &-2N\nabla_{(a}H_{b)}-\frac{1}{2}Nt^c t^d\Pi_{cd}\Pi_{ab}-Nt^c \Pi_{ci}g^{ij}\Phi_{jab}\notag\\
                      &+N\gamma_0\left(2\delta^c{}_{(a} t_{b)}-\psi_{ab}t^c\right)\left(H_c+\Gamma_c\right)\notag\\
                      &-\gamma_1\gamma_2N^i\Phi_{iab}\notag\\
                      &-16\pi N\left(T_{ab}-\frac{1}{2}\psi_{ab}T^c{}_c\right),\\
  \label{eq:metric_derivative_evolution}
  \partial_t\Phi_{iab}-&N^k\partial_k\Phi_{iab}+N\partial_i\Pi_{ab}-N\gamma_2\partial_i\psi_{ab}\notag\\
  =&\frac{1}{2}Nt^c t^d\Phi_{icd}\Pi_{ab}+Ng^{jk}t^c\Phi_{ijc}\Phi_{kab}\notag\\
                      &-N\gamma_2\Phi_{iab},
\end{align}
where $H_a$ is the so-called gauge source function and must satisfy the
constraint $H_a=\psi_{ab}\Gamma^b_{cd}\phi^{cd}$.  The parameters
$\gamma_0,\gamma_1$ and $\gamma_2$ are described in
$\S$\ref{sec:ConstraintDamping}. The first-order massless-Klein-Gordon system is

\begin{align}
  \label{eq:sw_psi_evolution}
  \partial_t\psi =& N^i\partial_i\psi-N\Pi+\gamma^{KG}_1N^i\left(\partial_i\psi-\Phi_i\right),\\
  \label{eq:sw_pi_evolution}
  \partial_t\Pi=&N\Pi K+N^i\partial_i\Pi+N\Phi_i g^{jk}\Gamma^i_{jk}\notag\\
                  &+\gamma^{KG}_1\gamma^{KG}_2N^i\left(\partial_i\psi-\Phi_i\right)\notag\\
                  &-g^{ij}\left(N\partial_j\Phi_i+\Phi_j\partial_i N\right),\\
  \label{eq:sw_phi_evolution}
  \partial_t\Phi_i=&-N\partial_i\Pi-\Pi\partial_i N-\gamma^{KG}_2N\left(\Phi_i-\partial_i\psi\right)\notag\\
                  &+N^j \partial_j\Phi_i + \Phi_j\partial_i N^j.
\end{align}
The parameters $\gamma^{KG}_1$ and $\gamma^{KG}_2$ are described in
$\S$\ref{sec:ConstraintDamping}, and $K$ is the trace of the extrinsic
curvature.

\section{Initial Data}\label{sec:InitialData}
We generate initial data for the evolutions by solving the extended conformal
thin-sandwich equations~\cite{Pfeiffer:2002iy} using the spectral elliptic
solver~\cite{2003CoPhC.152..253P} in \texttt{SpEC}~\cite{SpECwebsite}. The
contributions to the equations from the scalar field are given by
\begin{align}
  \label{eq:rhoID}
  \rho =&t^a{}t^b{}T_{ab}=\frac{1}{2}\left(\Pi{}^2+g^{ij}\Phi_i\Phi_j\right),\\
  \label{eq:momentumID}
  S^i =&-g^{ij}t^a{}T_{aj}=g^{ij}\Pi\Phi_{j},
\end{align}
and
\begin{align}
  \label{eq:stressID}
  S =& g_{ij}g^{ia}g^{jb}T_{ab}=\frac{1}{2}\left(3\Pi{}^2-g^{ij}\Phi_i\Phi_j\right),
\end{align}
where $g^{ia}$ projects the spacetime index $a$ onto the spatial hypersurface
orthogonal to $t^a$.

Let $r=\delta_{ij}x^ix^j$ and
\begin{align}
  \label{eq:1d gaussian}
  f(r) = \varphi_0 \exp\left[-\left(\frac{r-r_0}{\sigma}\right)^2\right].
\end{align}
For concreteness we focus on three types of initial data: spherically symmetric
data given by
\begin{align}
  \label{eq:Spherical ID}
  \varphi(t,x^i)=\varphi_{\mathrm{sph}}=\frac{f(-r)+f(r)}{r},
\end{align}
data where the second term has no $y$-coordinate dependence
(recall $xz\sim r\cos\phi\sin2\theta$) similar to that studied
in~\cite{Healy:2013xia,Clough:2016jmh}
\begin{align}
  \label{eq:Reflection ID}
  \varphi(t,x^i)=\varphi_{\Re (Y^2_1)}:=\varphi_{\mathrm{sph}}\left(1-\delta\cos\phi\sin2\theta\right),
\end{align}
and finally generic initial data of the form
\begin{align}
  \label{eq:Generic ID}
  \varphi(t,x^i)=\varphi_{3-d}:=\varphi_{\mathrm{sph}}
  &\left\{1-\frac{\delta}{1.56}\left[(\cos\phi+\sin\phi)\sin2\theta\right.\right.\notag\\
  &\left.\left.-\left(3\cos^2\theta-1\right)\right]\right\}.
\end{align}
The conjugate momentum to the $\varphi$ in the spherically symmetric case is
given by
\begin{align}
  \Pi_{\mathrm{sph}}=\frac{\partial_rf(-r)-\partial_rf(r)}{r},
\end{align}
and is multiplied by the same non-spherical terms as $\varphi$.  This is ingoing
spherical wave initial data.  The numerical factor $1.56$ is chosen so that when
$\delta=1$, the maximum of the second term is approximately unity. We choose
$\sigma=1$ and $r_0=5$ for the results presented here. For the initial
data~\eqref{eq:Reflection ID} we (arbitrarily) choose $\delta=0.9$ and for data
given by~\eqref{eq:Generic ID} we choose $\delta=1$.

\section{Numerical Methods}\label{sec:NumericalMethods}

\subsection{Domain Decomposition}
\texttt{SpEC} decomposes the computational domain into possibly overlapping
subdomains. Within each subdomain a suitable set of basis functions that depends
on the topology of the subdomain is chosen to approximate the solution. The
domain decomposition for finding the initial data is a cube at the center with
an overlapping spherical shell that is surrounded by concentric spherical
shells. For the evolution, a filled sphere surrounded by non-overlapping
spherical shells is used until a black hole forms. At this point a ringdown or
excision grid nearly identical to that used during the ringdown phase of binary
black hole merger evolutions is used~\cite{Scheel:2008rj, Szilagyi:2009qz,
  Hemberger:2012jz}. The ringdown grid consists of a set of non-overlapping
spherical shells with the inner shell's inner radius approximately $94\%$ of the
apparent horizon radius.

\subsection{Dual Frames and Mesh Refinement}
To resolve the large range of spatial and temporal scales required,
finite-difference codes typically use adaptive mesh refinement (AMR).  However,
for the spatiotemporal scales required here, AMR is computationally
prohibitively expensive in 3+1 dimensions without any symmetries.

\texttt{SpEC} achieves its high accuracy by using spectral methods to solve the
PDEs rather than finite differencing. In addition, two further tools are
employed to achieve high accuracy: dual frames~\cite{Scheel:2006gg,
  Szilagyi:2009qz, Hemberger:2012jz} and spectral AMR~\cite{Szilagyi:2014fna}.

In the dual frames approach, the PDEs are solved in what is called the grid
frame. This frame is related to the ``inertial frame'', the frame in which the
PDEs are originally written, by time-dependent spatial coordinate maps. The dual
frames method ``moves'' the grid points inward as the scalar field collapses,
which gives an additional two orders of magnitude of resolution compared to the
initial inertial coordinates without the use of any mesh refinement. We also
employ a coordinate map to slowly drift the outer boundary inward so that any
constraint-violating modes near the outer boundary are propagated out of the
computational domain. While the slow drift of the outer boundary is not
essential for stability, it is helpful in long evolutions.

Denote the coordinate map that moves the grid points inward during collapse by
$\mathcal{M}_{\mathrm{scaling}}$ and the map that drifts the outer boundary
inward by $\mathcal{M}_{\mathrm{drift}}$. Then the coordinate map used during
collapse before a black hole forms is given by
$\mathcal{M}_{\mathrm{collapse}}=\mathcal{M}_{\mathrm{drift}}\circ\mathcal{M}_{\mathrm{scaling}}$.
The mapping $\mathcal{M}_{\mathrm{collapse}}$ relates the initial coordinates,
$\bar{x}^i$ to the grid coordinates $x^i$ by
$\bar{x}^i=\mathcal{M}_{\mathrm{collapse}}x^i$. The specific spatial coordinate
map we use for both $\mathcal{M}_{\mathrm{drift}}$ and
$\mathcal{M}_{\mathrm{scaling}}$ is of the form
\begin{align}
  \label{eq:cubicScale}
  \bar{r}=a(t)r+\left[1-a(t)\right]\frac{r^3}{r_{\mathrm{outer}}^2},
\end{align}
where $r=\delta_{ij}x^ix^j$, $\bar{r}=\delta_{ij}\bar{x}^i\bar{x}^j$, $a(t)$ is
a time-dependent function we call an expansion factor, and $r_{\mathrm{outer}}$
is a parameter of the map. For $\mathcal{M}_{\mathrm{scaling}}$ we choose
\begin{align}
  \label{eq:aScaling}
  a_{\mathrm{scaling}}(t) = A
  \exp\left[-{\left(\frac{t}{\sigma_{\mathrm{scaling}}}\right)}^{2n}\right]
  +B
\end{align}
with $A=0.99$, $B=0.01$, $n=2$ and $\sigma_{\mathrm{scaling}}=3.8$. The value of
$r_{\mathrm{outer}}$ for $\mathcal{M}_{\mathrm{scaling}}$ is
$r_{\mathrm{outer}}=100$. For $\mathcal{M}_{\mathrm{drift}}$ we use
$r_{\mathrm{outer}}=180$ and
\begin{align}
  \label{eq:aDrift}
  a_{\mathrm{drift}}(t)=1+v\frac{t^3}{b+t^2},
\end{align}
with $b=10^{-4}$ and $v=-3.23\times10^{-3}$. We find these choices for the
coordinate maps lead to accurate and stable long-term evolutions with sufficient
resolution to resolve both scaling and echoing.

After an apparent horizon is found we switch over to an excision grid and use
the same coordinate maps used in the ringdown portion of the binary black hole
evolutions~\cite{Scheel:2008rj, Szilagyi:2009qz, Hemberger:2012jz}.
Specifically, we excise the interior of the apparent horizon with the excision
surface's radius being approximately 94 per cent of the apparent horizon's
coordinate radius. Near the apparent horizon, all the characteristics are
directed toward the center of the apparent horizon and so no boundary conditions
need to be imposed there. Thus, as long as the excision surface remains close to
the apparent horizon, the simulation remains stable without the need to impose
additional boundary conditions.  One difficulty is that during the very early
phase of ringdown the apparent horizon's coordinate radius increases very
rapidly. To deal with the rapid expansion, a control system is used to track the
apparent horizon and adjust the location of the excision boundary to follow the
apparent horizon~\cite{Scheel:2006gg, Scheel:2008rj, Hemberger:2012jz}.

While the spatial coordinate maps work extremely well for resolving the small
length scales that appear near the critical solution, they do not provide any
guarantees about the truncation error of the simulations. The temporal error is
controlled by using an adaptive, fifth-order Dormand-Prince time stepper. The
spatial error is controlled using the spectral AMR algorithm described
in~\cite{Szilagyi:2014fna}.  Using AMR we control the relative error in the
metric, the spatial derivative of the metric and the conjugate momentum of the
metric. For the results presented in this manuscript we set a relative maximum
spatiotemporal error of $10^{-8}$.

\subsection{Gauge Choice}
In binary black hole evolutions with the GH system, large constraint violations
occur unless an appropriate gauge condition is chosen. The key ingredient in a
successful choice~\cite{Lindblom:2009tu} is to control the growth of
$\sqrt{g}/N$, where $g$ is the determinant of the spatial metric.  As one might
expect, evolutions of critical behavior at black hole formation require even
more stringent control of the gauge than in binary simulations. We find that
without such control, explosive growth in both $\sqrt{g}/N$ and $1/N$ prevents
the code from finding an apparent horizon before the constraints blow up and the
evolution fails. Accordingly, we adopt a modified version of the damped harmonic
gauge used in Ref.~\cite{Lindblom:2009tu}:
\begin{align}
  \label{eq:targetGauge}
  H_a=&\left[\mu_{L,1}\log\left(\frac{\sqrt{g}}{N}\right)
        +\mu_{L,2}\log\left(\frac{1}{N}\right)\right]t_a\notag\\
      &-\mu_S N^{-1}g_{ai}N^i.
\end{align}
The coefficients $\mu_{L,1}$, $\mu_{L,2}$ and $\mu_{S}$ are described below.

Fortunately, the region of the spatial hypersurfaces where $\sqrt{g}/N$ diverges
is different from that where $1/N$ diverges and so having the coefficients
$\mu_{L,1}$ and $\mu_{L,2}$ depend on $\log(\sqrt{g}/N)$ and $\log{1/N}$
respectively allows us to control both divergences with a single equation.  The
functional forms of the coefficients are
\begin{align}
  \label{eq:muL1}
  \mu_{L,1}=&R(t)W(x^i)\left[\log\left(\frac{\sqrt{g}}{N}\right)\right]^4,\\
  \label{eq:muL2}
  \mu_{L,2}=&R(t)W(x^i)\left[\log\left(\frac{1}{N}\right)\right]^4,
\end{align}
and
\begin{align}
  \label{eq:muS}
  \mu_{S}=&\mu_{L,1}.
\end{align}
The roll-on function $R(t)$ is given by
\begin{align}
  \label{eq:rollon}
  R(t)=1-\exp\left[-\left(\frac{t-t_0}{\sigma_t}\right)^4\right],
\end{align}
where we choose $t_0=0$ and $\sigma_t=2$, while the spatial weight function,
$W(x^i)$ is given by
\begin{align}
  \label{eq:spatialWeight}
  W(x^i)=\exp\left[-34.54\left(\frac{r}{r_{\max}}\right)^2\right],
\end{align}
where we set $r_{\max}=30$. The function $R(t)$ is used to transition from the
initial maximal slicing to the damped harmonic gauge needed later in the
evolution, while $W(x^i)$ makes the gauge be pure harmonic near the outer
boundary of the computational domain. The $\log$ factors in Eq.~\eqref{eq:muL1}
and~\eqref{eq:muL2} make the gauge pure harmonic in the region of the spatial
slice where $\sqrt{g}/N$ and $1/N$ are near unity, respectively. We found that
using the fourth power as opposed to the second power that is typically used for
controlling the growth of $\sqrt{g}/N$ in binary black hole evolutions is
required for stable long-term evolutions.

\subsection{Constraint Damping}\label{sec:ConstraintDamping}

Both the Klein-Gordon and the GH system have constraints that must remain
satisfied during evolutions. For the Klein-Gordon system the constraint is
\begin{align}
  \label{eq:KgConstraint}
  \mathcal{C}^{KG}_i=\partial_{i}\psi-\Phi_{i}=0.
\end{align}
The constraints for the GH system are given in reference~\cite{Lindblom:2005qh}.

Failure to satisfy the constraints indicates that the numerical simulation is no
longer solving the physical system of interest and should not be trusted. To
control growth of constraint violations from numerical inaccuracies, constraint
damping parameters are added to the evolution equations. For the GH system the
constraint damping parameters are $\gamma_0, \gamma_1$ and $\gamma_2$, and for
the Klein-Gordon system $\gamma_1^{\mathrm{KG}}$ and
$\gamma_2^{\mathrm{KG}}$. See
Eqs.(\ref{eq:metric_evolution}--\ref{eq:sw_phi_evolution}) for how the
constraint damping parameters appear in the evolution equations. We find that
choosing $\gamma_1^{\mathrm{KG}}=1$ and $\gamma_2^{\mathrm{KG}}=0$ works well
for the scalar field. For the GH system, finding good constraint damping
parameters is more difficult, especially during ringdown. The dimensions of the
constraint damping parameters are $\mathrm{time}^{-1}$, which suggests that for
smaller black holes where the characteristic time scale is shorter, the
constraint damping parameters must be increased. During ringdown we choose
\begin{align}
  \gamma_0 &= A_0\exp\left(-\frac{r^2}{10^2}\right)+10^{-3},\\
  \gamma_1 &= A_1\left[\exp\left(-\frac{r^2}{1000^2}\right)-1\right],\\
  \gamma_2 &= A_2\exp\left(-\frac{r^2}{10^2}\right)+10^{-3},
\end{align}
with $A_0\in [20, 100]$, $A_1=0.999$, and $A_2\in[20, 80]$. Larger values of
$A_0$ and $A_2$ are used for smaller black holes.  During the collapse phase of
the evolutions we find less sensitivity to the choice of the damping
parameters. We use the same functional form as during the ringdown but always
choose $A_0 = A_2 = 20$.

\section{Results}\label{sec:Results}

All files used to produce figures in this paper, including the data, are
available from the arXiv version of this paper.

\subsection{Scaling}\label{sec:Scaling}
In this section we present two sets of scaling relations.  The first involves
the final mass of the black hole $M_{\mathrm{BH}}$ for supercritical
evolutions. For each class of initial data we evolve the data with amplitudes
large enough that a black hole forms and gradually decrease the amplitude. While
decreasing the amplitude we focus on simulations that form a black hole. Rather
than performing a binary search to estimate $p_\star$, we fit the relationship
$\ln(M_{\mathrm{BH}})=\gamma\ln(p/p_\star-1)+C$ to the data for $\gamma$,
$p_\star$, and $C$, where we take $p$ to be the amplitude $\varphi_0$ of the
initial data. We then use the $p_\star$ from the fit to determine an amplitude
that should form a black hole but is closer to the critical solution. This is
repeated until $\log_{10}(p/p_\star-1)\approx-6$, the target value. Choosing
suitable values of $p$ to fit for $\gamma$ and $\Delta$ is tricky. We describe
our procedure in the \hyperref[sec:Appendix]{Appendix}. Note that the
relationship used for determining which amplitude to use next is not used for
analyzing the results.

The second scaling relation involves, $R_{\max}$ the maximum Ricci scalar at the
center for subcritical evolutions.  We run simulations to obtain an
approximately even distribution of masses and maximum Ricci scalars for
$\ln(p/p_\star-1)\in(-14,-5]$. We estimate the errors in the final mass of the
black hole and $R_{\max}$ using convergence tests with values of $p$ nearest
$p_\star$.

Once we have reached the target number of simulations, with the lowest
amplitude that forms a black hole having $\log_{10}(p/p_\star-1)\approx-6$, we
fit the mass of the resulting black hole to
\begin{align}
  \label{eq:SineMassFit}
  \ln(M_{\mathrm{BH}})=&\gamma^M\ln(p/p_\star-1)+C^M\notag\\
                       &+A^M\sin\left[w^M\ln(p/p_\star-1)+\delta^M\right],
\end{align}
as suggested in~\cite{Gundlach:1996eg, Hod:1996az}. Note that the superscript
$M$ is not an exponent but denotes that parameter was obtained from fitting to
the mass of the black hole rather than the maximum Ricci scalar at the
center. We find that the probability of $\chi^2$ and the reduced $\chi^2$ are
better for this function than the one where the sinusoidal term is omitted. We
fit for all parameters in~\eqref{eq:SineMassFit}, including $p_\star$. The
fitting function used for the maximum Ricci scalar at the origin is
\begin{align}
  \label{eq:SineRicciFit}
  \ln(R_{\max})=&2\gamma^R\ln(p/p_\star-1)+C^R\notag\\
                &+A^R\sin\left[w^R\ln(p/p_\star-1)+\delta^R\right].
\end{align}
However, for consistency we use the value of $p_\star$ obtained from fitting to
the masses when fitting to the maximum Ricci scalar as well.

\begin{figure}[]
  \centering \includegraphics[width=0.47\textwidth]{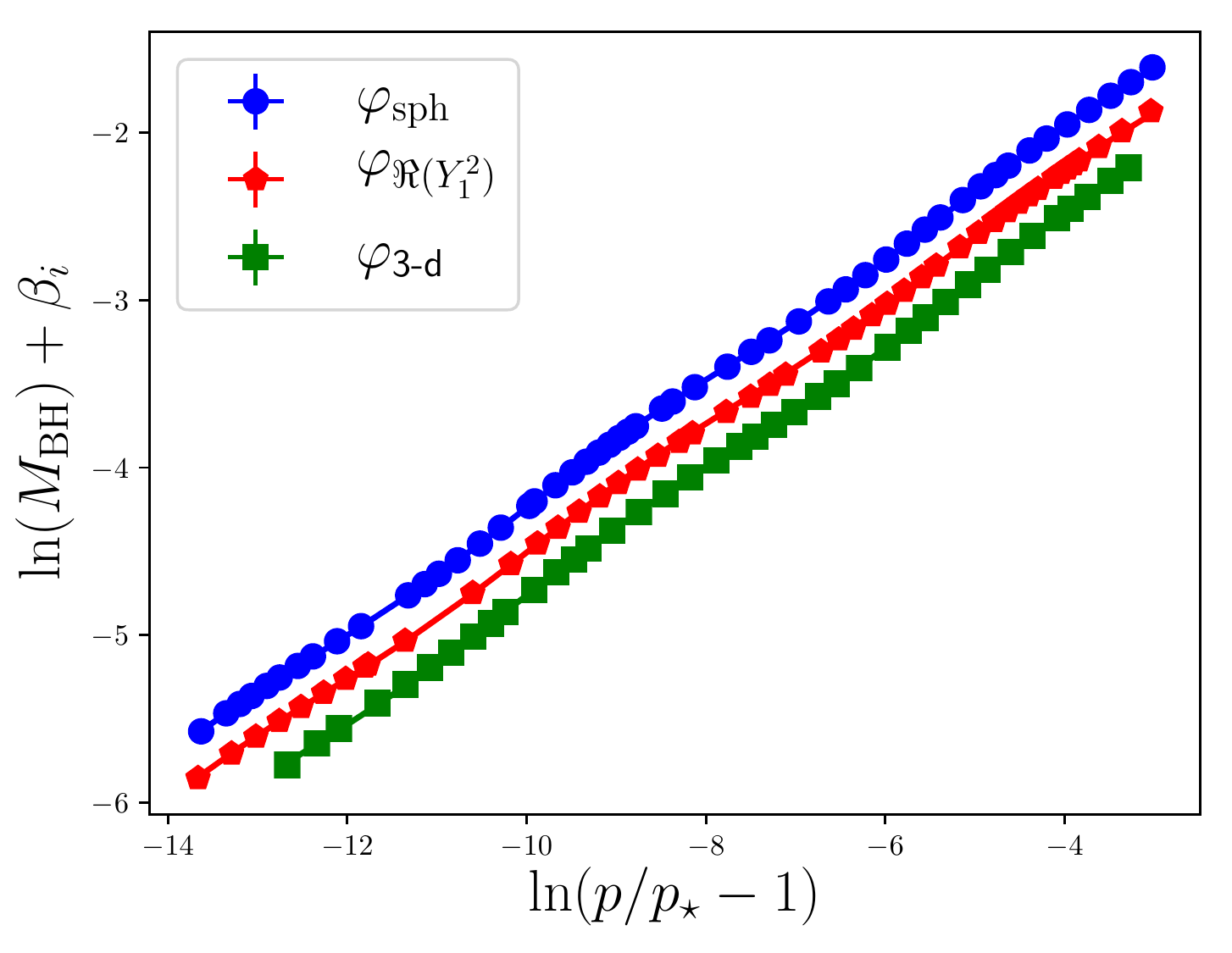}
  \caption{Plotted is $\ln(M_{\mathrm{BH}})$ as a function of $\ln(p/p_\star-1)$
    for the three types of initial data studied. We find critical exponents
    \gammamass{}. We shift the curves vertically by \betamass{} so that data
    points from different initial data are easily
    distinguished.}\label{fig:ScalingMasses}
\end{figure}

In Fig.~\ref{fig:ScalingMasses} we plot $\ln(M_{\mathrm{BH}})$ as a function of
$\ln(p/p_\star-1)$ for the three types of initial data studied. For data
$\varphi_{\Re(Y^2_1)}$ we arbitrarily choose $\delta=0.9$, which is a large
deviation from the spherical solution. For reference, when $\delta=1$ the scalar
field profile is zero at the zeros of $1-\cos(\varphi)\sin(2\theta)$. For
initial data $\varphi_{\text{3-d}}$ we choose $\delta=1$, an even stronger
deviation from spherical symmetry. In Fig.~\ref{fig:ScalingMasses} we offset the
curves vertically by \betamass{} so that they do not overlap and are easier to
compare. The critical exponents we find are \gammamass{}, where the number in
parentheses is the uncertainty in the last digit. These are all close to the
accepted value for spherically symmetric initial data,
$0.374\pm0.001$~\cite{Gundlach:1996eg} strongly suggesting that the spherical
mode dominates.

\begin{figure}[]
  \centering \includegraphics[width=0.47\textwidth]{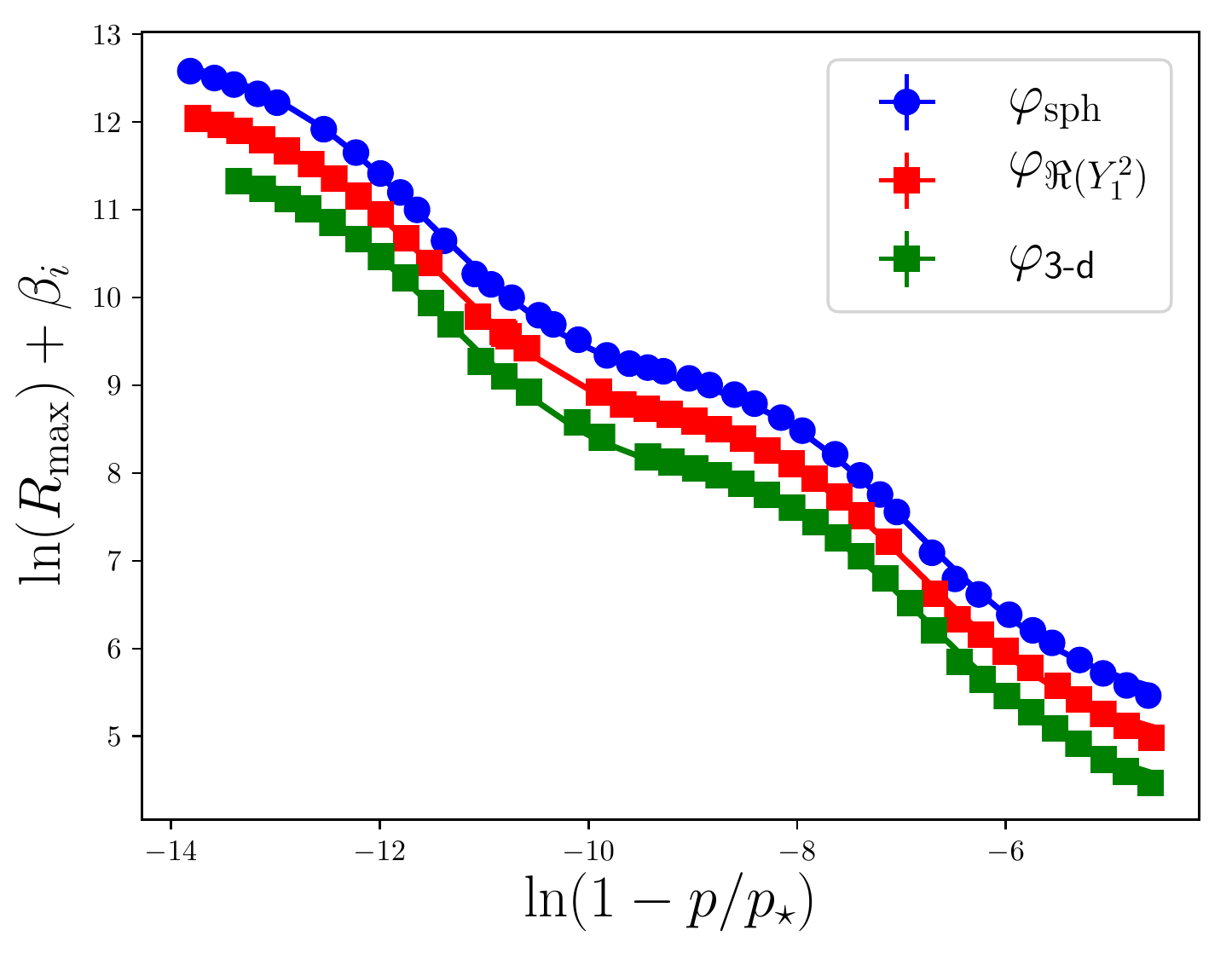}
  \caption{Plotted is $\ln(R_{\max})$ as a function of $\ln(1-p/p_\star)$ for
    the three types of initial data studied. We find critical exponents
    \gammaricci{}. We shift the curves vertically by \betaricci{} so that data
    points from different initial data are easily
    distinguished.}\label{fig:ScalingRicci}
\end{figure}

In addition to studying the final mass of the resulting black hole, we
follow~\cite{Garfinkle:1998va} and calculate the maximum Ricci scalar at the
center of the collapse for subcritical evolutions.  In
Fig.~\ref{fig:ScalingRicci} we plot $\ln(R_{\max})$ as a function of
$\ln(p/p_\star-1)$ along with a fit using Eq.~(\ref{eq:SineRicciFit}) for the
initial data studied. We again offset the plots vertically by amounts
\betaricci{} to aid readability. In this case we find critical exponents
\gammaricci{}, which are comparable to the values for mass scaling and to the
accepted value in spherically symmetric critical collapse,
$\gamma=0.374\pm0.001$.

\subsection{Echoing}\label{sec:Echoing}

\begin{figure}[]
  \centering
  \includegraphics[width=0.47\textwidth]{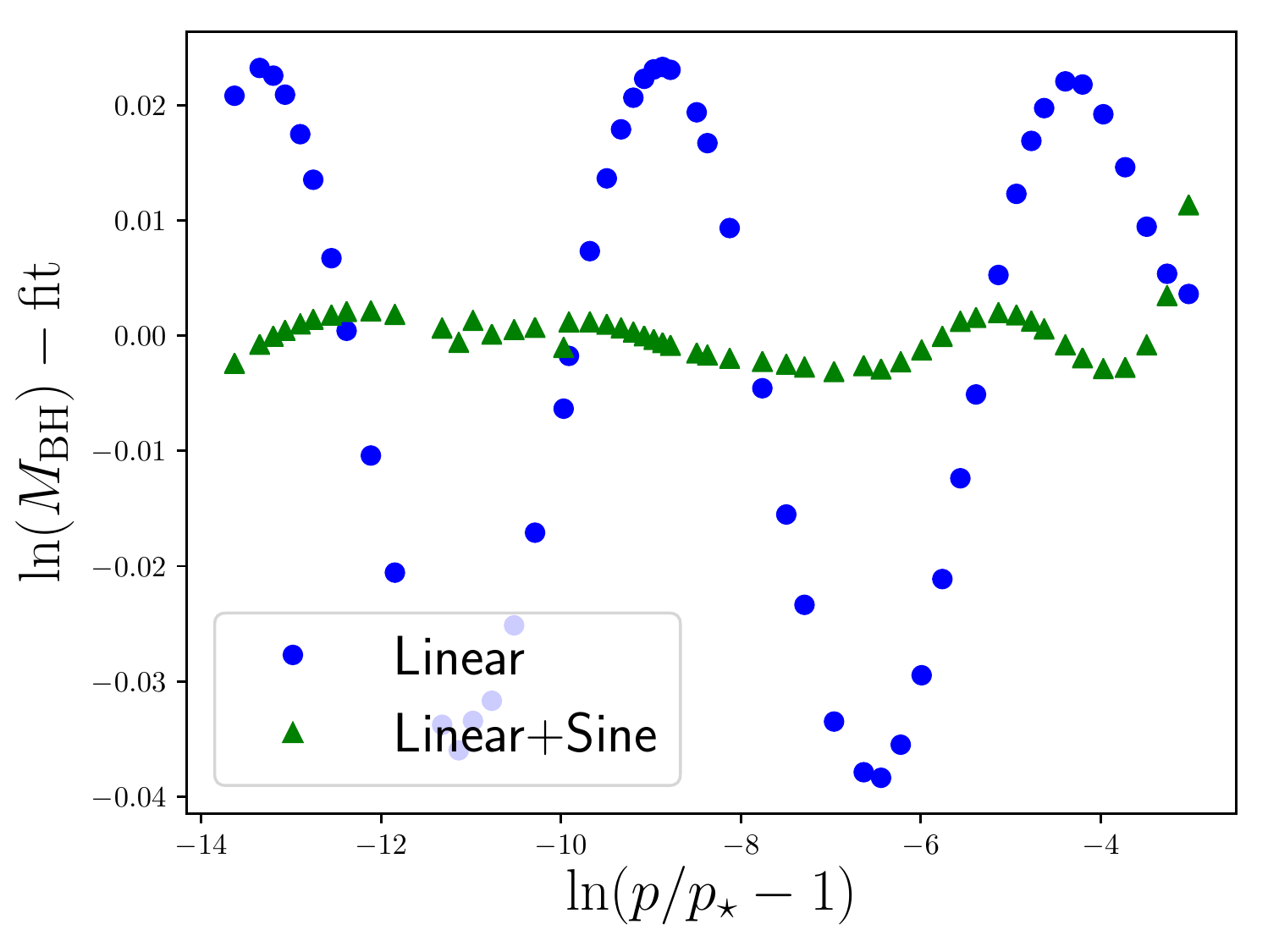}
  \caption{The residuals of the fitting
    $\ln(M_{\mathrm{BH}})=\gamma^M\ln(p/p_\star-1)+C$ (blue dots)
    and Eq.~\ref{eq:SineMassFit} (green triangles) to the black hole
    masses for the spherical symmetry case, $\varphi_{\rm{sph}}$. The
    sinusoidal residual of the straight line fit is identical to what
    is observed
    in~\cite{Hod:1996az}.}\label{fig:ResidualsSs}
\end{figure}

Having studied the scaling we now turn to the fine
structure and echoing of the critical behavior. Echoing of any gauge-invariant
quantity was described by Eq.~\eqref{eq:rescaling} above. A small-amplitude
sinusoidal  modulation about the straight line expected from critical
behavior was conjectured and observed
in~\cite{Hod:1996az}. Fig.~\ref{fig:ScalingMasses} and
~\ref{fig:ScalingRicci} both show this feature. In
Fig.~\ref{fig:ResidualsSs} we plot the residuals when fitting only the
linear term and when fitting the linear plus sine term for the
spherically symmetric mass scaling case.\footnote{The residuals of the
  fits for non-spherical initial data and for Ricci scaling are qualitatively
identical.} The sinusoidal modulation is
much clearer in Fig.~\ref{fig:ResidualsSs} than in
Fig.~\ref{fig:ScalingMasses}.

From the fit, Eq.~(\ref{eq:SineMassFit}), we estimate the period,
$T=2\pi/w$. In~\cite{Hod:1996az} it was found that the
relationship between the echoing period, $\Delta$ and the scaling
period, $T$ is $T=\Delta/ (2 \gamma)$. To test this relationship, we calculate
$\Delta$ using $T$ and also by estimating it directly from the
Ricci scalar at the origin as a function of the logarithmic time,
$-\ln(1-\tau/\tau_\star)$.
$\tau$ is the proper time at the origin given by
\begin{align}
  \label{eq:ProperTimeOrigin}
  \tau=\int_0^t N(\tilde{t}, 0)d\tilde{t},
\end{align}
and $\tau_\star$ is the accumulation time of the self-similar solution.

\begin{figure}[]
  \centering
  \includegraphics[width=0.47\textwidth]{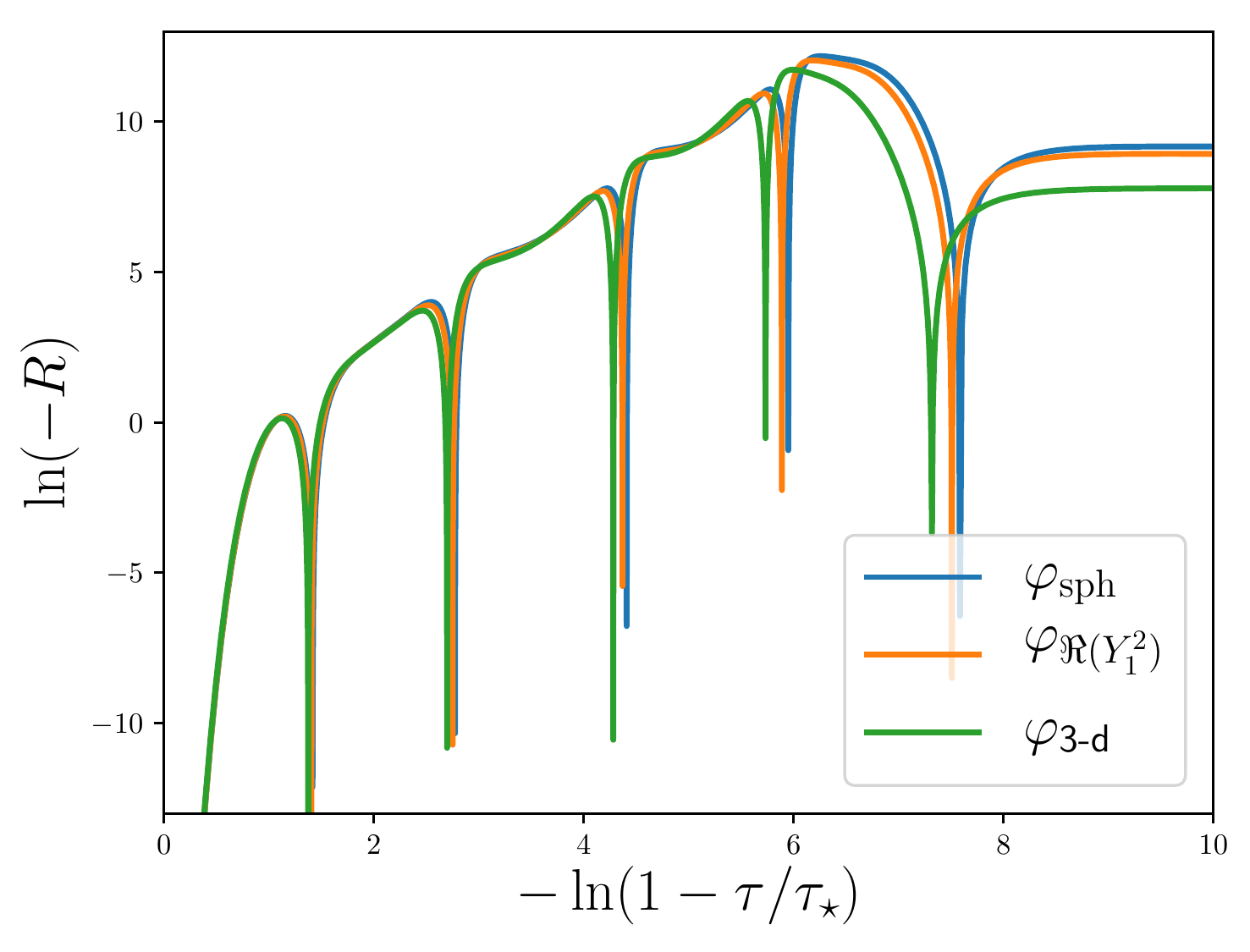}
  \caption{Plotted is $\ln(R(t,r=0))$ as a function of
    $\ln(1-\tau/\tau_\star)$ for the three types of initial data
    studied. The echoing is clearly visible and very similar between
    the different evolutions, which all have
    $\ln(1-p/p_\star)\approx-6$. The echoing period is
    $\Delta=3.2\pm0.1$ for all
    simulations.}\label{fig:Echoing}
\end{figure}

We find that despite being able to resolve the fine
structure and knowing $p_\star$ to six significant figures, the
estimate of $\tau_\star$ from the apparent horizon formation time is
only accurate to about two digits.  This is because the
formation time of an apparent horizon is a gauge-dependent quantity. We
estimate
$\tau_\star$ by assuming that the logarithmic time between
successive echoes becomes constant and adjusting $\tau_\star$ until
this is true. The resulting $\tau_\star$ is consistent with what we estimate
from apparent horizon formation times. In Fig.~\ref{fig:Echoing} we
plot $\ln(R(t,r=0))$, a
geometric invariant, which shows the expected echoing that has been
studied in previous work~\cite{Garfinkle:1998va,Sorkin:2005vz}. From
Fig.~\ref{fig:Echoing} we estimate the echoing period to be
$\Delta=3.2\pm0.1$.

\begin{table}
  \centering
  \begin{tabularx}{\columnwidth}{@{\extracolsep{\stretch{1}}}*{7}{c}@{}}
    \hline
    Initial Data & $2\gamma^MT^M$ &
    $2\gamma^RT^R$ & $\Delta_{\mathrm{echoing}}$ \\ \hline
    $\varphi_{\mathrm{sph}}$ & $3.46\pm0.01$ & $3.557\pm0.001$ &
    $3.2\pm0.1$ \\
    $\varphi_{\Re (Y^2_1)}$ & $3.46\pm0.02$ & $3.518\pm0.002$ & $3.2\pm0.1$ \\
    $\varphi_{\mathrm{3-d}}$ & $3.67\pm0.04$ & $3.512\pm0.003$ & $3.2\pm0.1$ \\ \hline
  \end{tabularx}
  \caption{Comparison of $2\gamma^MT^M$ and the echoing period $\Delta$.
    In~~\cite{Hod:1996az} it was found that $\Delta=2\gamma T$, which
    we are unable to verify within our error estimates.
    The accepted value of the echoing period in spherical symmetry  is
    $\Delta=3.4453\pm0.0005$~\cite{Gundlach:1996eg}.}\label{tab:EchoingPeriods}
\end{table}

In Table~\ref{tab:EchoingPeriods} we summarize and compare direct
estimates of $\Delta$ to $2\gamma T$. Specifically, we find that
$2\gamma^MT^M\approx3.46$, near the best known value of
$\Delta=3.4453\pm0.0005$~\cite{Gundlach:1996eg}. For simulations
that do not form a horizon, where we compute $2\gamma^RT^R$ from the
Ricci scalar scaling plot, Fig.~\ref{fig:ScalingRicci}, we find
that
$2\gamma^R_{\mathrm{sph}}T^R_{\mathrm{sph}}=3.556\pm0.001$,
$2\gamma^R_{\Re  (Y^2_1)}T^R_{\Re  (Y^2_1)}=3.518\pm0.002$, and
$2\gamma^R_{\text{3-d}}T^R_{\text{3-d}}=3.512\pm0.003$. The discrepancy
between $2\gamma T$ from mass scaling and Ricci scalar scaling is
currently not understood. When studying the echoing of $\ln(-R(t,r=0))$, we find
$\Delta=3.2\pm0.1$, where the larger error is explained by the difficulty
in estimating $\tau_\star$.

\begin{figure}[]
  \centering
  \includegraphics[width=0.47\textwidth]{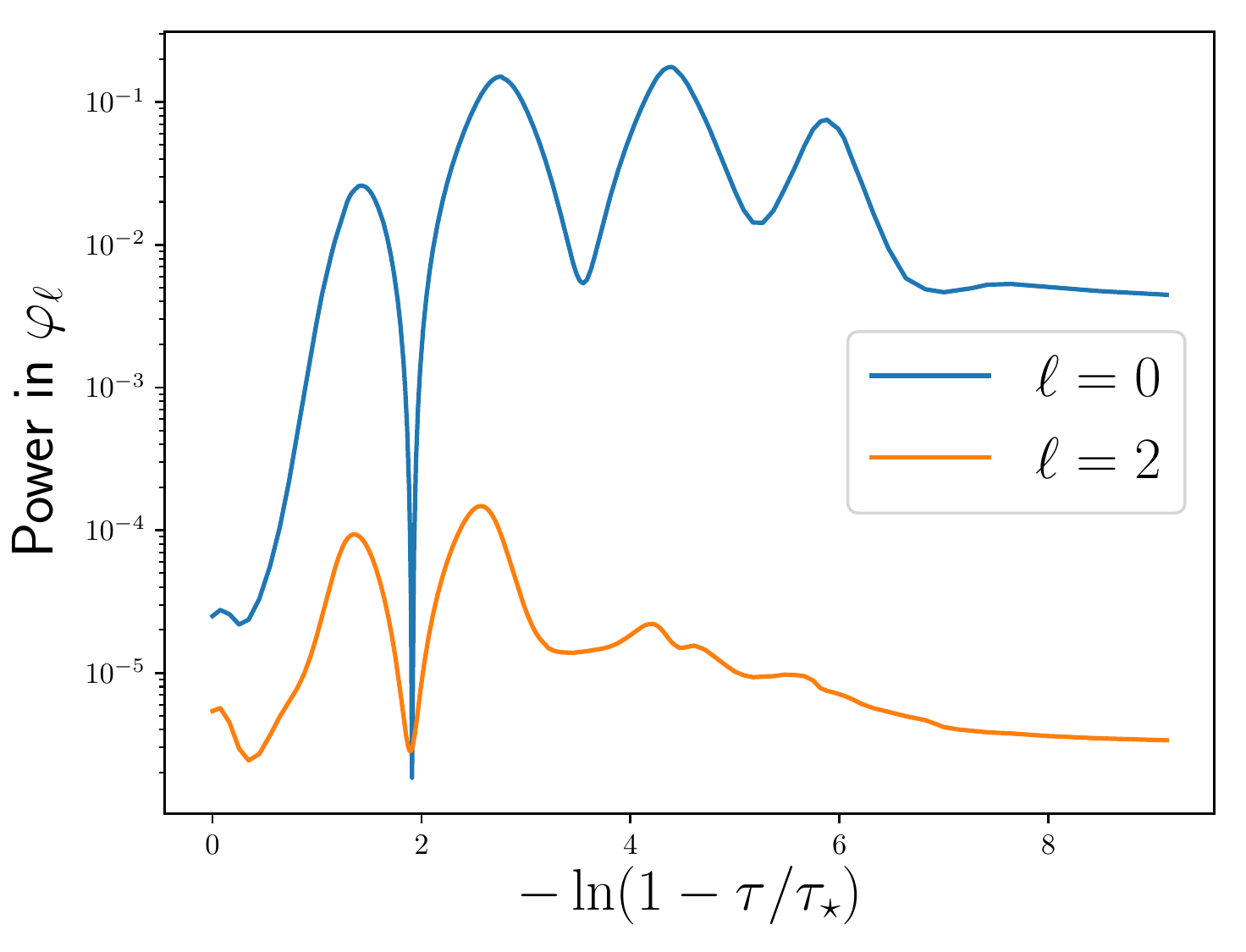}
  \caption{The power in $\varphi_\ell$ for $\ell=0, 2$ for the $\Re(Y^2_1)$
    initial data with $\varphi_0=0.07586803$.}\label{fig:PowerPsiPlanar}
\end{figure}

A power spectrum analysis shows that the spherical mode dominates the
evolution. We define the power in a given $\ell$-mode as
\begin{align}
  P_\ell = \frac{1}{N_r}
  \sum_{i=0}^{N_r-1}\sum_{m=-\ell}^{\ell}\left|C_{i,\ell,m}\right|^2
\end{align}
where $N_r$ is the number of radial points, and $C_{i,\ell,m}$ are the
coefficients in spectral expansion. This definition is consistent with
Parseval's theorem given that
\begin{align}
  \int \rvert Y_m^\ell(\theta, \phi)\rvert^2d\Omega=1.
\end{align}
Also note that with this definition at a given radius
\begin{align}
  \int \rvert f(\theta, \phi)\rvert^2d\Omega = \sum_{\ell=0}^{\infty}P_\ell.
\end{align}

For the $\Re(Y^2_1)$ data we find that initially
\begin{align}
  \frac{P_2}{P_0} = \frac{27}{125}
  \Rightarrow \frac{P_2}{\sum_{\ell}P_\ell} = \frac{P_2}{P_0 + P_2} \approx 0.18,
\end{align}
or that approximately 18 percent of the power is in the $\ell=2$ mode. For the
3-d initial data we find that initially
\begin{align}
  \frac{P_2}{P_0} \approx 0.548
  \Rightarrow \frac{P_2}{\sum_{\ell}P_\ell} = \frac{P_2}{P_0 + P_2} \approx 0.35,
\end{align}
or that approximately 35 percent of the power is in the $\ell=2$ mode.

In Fig.~\ref{fig:PowerPsiPlanar} we plot the power in $\varphi_\ell$ for
$\ell=0, 2$ for the $\Re(Y^2_1)$ initial data. Fig.~\ref{fig:PowerPsiPlanar}
shows that the $\ell=2$ mode decays much more rapidly than the $\ell=0$ mode,
suggesting that the spherically symmetric critical solution is approached.
However, given the different initial data and that we are further from the
critical solution than~\cite{Choptuik:2003ac}, we are unable to corroborate or
dispute their results.

The initial data used in~\cite{Baumgarte:2018fev} is given by
\begin{align}
  \label{eq:Baumgarte data}
  \varphi_{Y_2^{2}} =& \varphi_0
  \exp\left(-\frac{r}{r_0}\right)\left[\sin^2\theta
                       +\left(1-\delta^2\right)\cos^2\theta\right]\notag \\
  =&\varphi_0
  \exp\left(-\frac{r}{r_0}\right)\left(1 - \delta^2 +
     \delta^2\sin^2\theta\right).
\end{align}
The deformation in this case is proportional to the $Y_2^{\pm2}$ spherical
harmonics as opposed to the $Y_2^1$ spherical
harmonic. Ref.~\cite{Baumgarte:2018fev} found that
for $\delta=0.75$ the critical behavior differs significantly from that of the
spherically symmetric evolutions. For example, the critical exponent is observed
to be $\gamma\approx0.306$. The percentage of the power
in the $\ell=2$ mode for $\delta=0.75$ is approximately 47 percent. This
is 12 percent more than our 3-d initial data that has behavior consistent
with the spherically symmetric evolutions. This raises the question as to
whether the reason~\cite{Baumgarte:2018fev} see different behavior is because of
the increased power in the $\ell=2$ modes or because the initial data is
proportional to the $Y_2^{\pm2}$ spherical harmonics instead of the $Y_2^1$
spherical harmonic. Work is underway to attempt to resolve this question.

\section{Conclusions}\label{sec:Conclusions}

We present results of a study of critical behavior in the 3-d gravitational
collapse of a massless scalar field with no symmetry assumptions. We are able to
resolve the dominant critical behavior as well as the fine structure in both
supercritical and subcritical evolutions. We use the Spectral Einstein Code,
\texttt{SpEC}~\cite{SpECwebsite} to perform the evolutions, with several key
changes to the gauge condition and constraint damping.  We study how the
critical exponent and echoing period obtained from the data depend on how close
to the critical solution the simulations are, as well as how the simulations are
distributed in parameter space. This is especially important in 3-d where
simulations are costly to perform.  We find the critical exponents to be
\gammamass{}, consistent with the accepted result in spherical symmetry of
$0.374\pm0.001$~\cite{Gundlach:1996eg}.  The accepted value of the echoing
period $\Delta$ in spherical symmetry is
$\Delta=3.4453\pm0.0005$~\cite{Gundlach:1996eg}, while we find echoing periods
$\Delta=3.2\pm0.1$ for all initial data consider. The discrepancy can be
attributed to the difficulty in directly measuring the echoing period.  We also
test the predicted relationship~\cite{Gundlach:1996eg, Hod:1996az} between the
echoing period and the fine structure of the scaling, $2\gamma T=\Delta$. We
find that for mass scaling \Deltamass{}, where $T^M$ is the period of the
sinusoidal fine structure.

The agreement of the critical exponent, echoing
period, and fine structure between the spherically symmetric and highly
non-spherical simulations leads us to conclude that even for initial data far
from spherical symmetry the critical solution is that of spherical symmetry.
However, the reason why our results differ from those of~\cite{Choptuik:2003ac}
and~\cite{Baumgarte:2018fev}, where data far
from spherical symmetry approaches a different critical solution, is not yet
fully understood. One reason for the discrepancy could be that in our data
approximately 18 percent of the total power is in the $\ell=2$ mode for the
$\Re(Y_1^2)$ initial data and 35 percent for the $3-d$ initial data, while
in~\cite{Baumgarte:2018fev} approximately 47 percent of the power is in the
$\ell=2$ mode. In other words, more power than we used is needed in the $\ell=2$
mode. Another possible reason is that~\cite{Baumgarte:2018fev} studied
$\ell=2, m=2$ initial data while we study $\ell=2,m=1$ initial data. Work is
underway to understand if either of these scenarios are responsible for the
discrepancy and to independently reproduce the simulations
of~\cite{Baumgarte:2018fev}.

\section{Acknowledgements}

We are grateful to Andy Bohn, Fran\c{c}ois H\'{e}bert, and Leo Stein for
insightful discussions and feedback on earlier versions
of this paper. We are also grateful to the anonymous referee for the
feedback. This work was supported in part by a Natural Sciences and Engineering
Research Council of Canada PGS-D grant to ND,
NSF Grant PHY-1606654 at Cornell University, and by a grant from the Sherman
Fairchild Foundation. Computations were performed
on the Zwicky cluster at Caltech, supported by the Sherman
Fairchild Foundation and by NSF award PHY-0960291.

\appendix*

\section{Choosing parameter values for simulations}\label{sec:Appendix}

When estimating the error in the critical exponent $\gamma$ and
$2\gamma T$, we find it important to not
only consider the error obtained from convergence tests, but also
to study how $\gamma$ and $2\gamma T$ depend on the number of data points, and
how close to $p_\star$ the data points are. The former should be thought of as
whether or not the $\ln(p-p_\star)$ space is sampled densely enough by the
simulations. While reducing this error
requires more (potentially costly) simulations, these simulations will be
similar in their dynamics to simulations that have already been performed and so
no algorithmic changes to the code are generally required. Determining how the
closeness to $p_\star$ affects $\gamma$ and $2\gamma T$ is a closely related,
but separate issue. We study both of these sources of errors separately, while
error estimates from convergence tests are included as error bounds on
$M_{\textrm{BH}}$ in the fits.

We use two methods to estimate the errors from our sampling of the
$\ln(p-p_\star)$ space. First, we use bootstrapping
to study how choosing different data points from within the datasets alters the
critical exponent and $2\gamma T$.
Second, we build a minimal grid that achieves the desired
error tolerances by using a greedy algorithm. If the minimal grid is the same as
or quite close to our grid we deduce that our grid may not be sufficiently dense
to accurately extract $\gamma$ and $2\gamma T$. We will now outline these
methods in more detail.

The goal of bootstrapping is to resample the dataset randomly to obtain
knowledge about how well the dataset represents the full population.
This is done by randomly selecting as many points as there are in the dataset,
while allowing repetition.
Eq.~\eqref{eq:SineMassFit} is then fit to the randomly selected points
to obtain the critical exponent and $2\gamma T$. By repeating this
procedure many times (we choose 10,000 times) we are able to plot a histogram
of the critical exponents and values of $2\gamma T$. The variance in both
$\gamma$ and $2\gamma T$ is then obtained by fitting a Gaussian to the
histograms.

Using bootstrapping we find that the critical exponents obtained from mass
scaling are left unchanged to within error with values \gammamassbootstrap{}. For
Ricci scaling, we find that the critical exponents also do not change within
error, but the error estimate from bootstrapping is larger by approximately
an order of magnitude than from the fit to the full dataset. The values
obtained for $\gamma$ from Ricci scaling are \gammariccibootstrap{}.
For $2\gamma T$ we find qualitatively similar results to the critical
exponent. Using data points from mass scaling we find \Deltamassbootstrap{}
and from Ricci scaling we find \Deltariccibootstrap{}.

A greedy algorithm is designed to find the approximate global minimum of a
problem by selecting the path that is a local minimum at each node in the
decision tree. In this case we
seek the optimal values of $p$ to determine $\gamma$ and $2\gamma T$. Assume
we have a minimal dataset that allows the fitting procedure to succeed. Then
the greedy algorithm randomly selects a new value of $p$ and computes the
corresponding black hole mass using Eq.\eqref{eq:SineMassFit}. If adding
the computed black hole mass to the dataset decreases the
error it is added, otherwise a new value of $p$ is selected and added if it
decreases the error in $\gamma$ and $2\gamma T$. This is repeated until the
error in $\gamma$ and $2\gamma T$ is below some specified tolerance.

The greedy algorithm method takes as input a range of $\ln(p/p_\star - 1)$
in which to sample points, as well as the fit parameters obtained from a
numerical study, i.e.~$p_\star, \gamma, C, A, w$, and $\delta$. Fake black
hole masses are computed using~\eqref{eq:SineMassFit} and adding a random
offset of at most $\pm10^{-3}$ to simulate numerical errors that would be
present in the numerical simulations. The algorithm initially randomly chooses
five (or six if also fitting for $p_\star$) data points on the specified
interval of $\ln(p/p_\star - 1)$. Next, points
are randomly added until the fitting algorithm successfully identifies fit
parameters. Then data points are randomly chosen and added to the dataset only
if they reduce $|\gamma_{\mathrm{greedy}}-\gamma_{\mathrm{simulation}}|$. Data
points are added until
$|\gamma_{\mathrm{greedy}}-\gamma_{\mathrm{simulation}}| < 10^{-4}$.

Using the greedy algorithm, we find that for
$\ln(p/p_\star - 1) \in [-14, -3]$ roughly 11 evenly spaced data points are
necessary to achieve the desired tolerance and for
$\ln(p/p_\star - 1) \in [-7, -3]$ approximately 15 evenly spaced data points
are necessary. This is far fewer than the roughly 40 to 50 data points used
for the fits to the numerical simulations. One reason for the difference
in the number of data points is that, as
indicated by the greedy algorithm results, initially when we do not know
$p_\star$ very accurately a denser grid is necessary to obtain a fit of
decent accuracy. Another reason is that after finding $p_\star$ to some accuracy
we preformed simulations to fill a grid with spacing of approximately 0.1
in $\log_{10}(p/p_\star-1)$, which, in hindsight was unnecessary.
Finally, a factor that was not accounted for in the greedy algorithm is that
the fitting algorithm may not succeed because a good initial guess for the
fit parameters is not known. The greedy algorithm always used the input that
we modeled the data from.

To estimate the errors in $\gamma$ and $2\gamma T$ arising from how far from
criticality the simulations are, we fit to only the lower or upper 25, 50 and 75
percent of data points. This provides insight into how many digits of the critical
amplitude $p_\star$ need to be resolved for the fits to be reliable. We
note that this test only determines whether or not $\gamma$ and $2\gamma T$
are locally constant in $\ln(p-p_\star)$ space. The test cannot make any
pdefinitive statements about $\gamma$ and $2\gamma T$ far outside this range,
though this remains true regardless of how close to machine precision
$1-p_\star/p$ is.

By fitting to only a subset of the dataset, we observe that when fewer than
two to three significant figures of $p_\star$ are known, the linear + sine fit
either fails to converge or else exhibits high sensitivity to the
initial guess of the fitting parameters.
However, the linear fit is still robust in this regime. Ultimately, we find
that knowing $p_\star$ to five or more significant figures provides
robust fit results and good accuracy of the local critical exponent and $2\gamma
T$, while knowing $p_\star$ to fewer digits can lead to convergent fits that are
biased by not having sufficiently resolved the sinusoidal oscillation.

\bibliography{refs}
\end{document}